\documentstyle[12pt]{article}
\def\be{\begin{equation}}
\def\ee{\end{equation}}
\def\re#1{({\ref{#1}})}
\def\ci#1{{\cite{#1}}}

\def\Tr{{\rm Tr}}

\def\p{{\partial}}
\def\pb{{\bar\partial}}
\def\a{{\alpha}}
\def\b{{\beta}}
\def\d{{\delta}}

\def\m{{\mu}}
\def\n{{\nu}}
\def\l{{\lambda}}
\def\s{{\sigma}}
\def\t{{\theta}}
\def\o{{\omega}}

\def\CC{{\cal C}}

\def\CE{{\cal E}}
\def\CF{{\cal F}}

\def\CM{{\cal M}}
\def\CN{{\cal N}}
\def\CO{{\cal O}}

\def\CS{{\cal S}}

\def\CU{{\cal U}}
\def\CV{{\cal V}}
\def\CX{{\cal X}}

\def\IC{\bf C}

\def\IH{\bf H}

\def\IP{\bf P}
\def\IR{\bf R}

\def\IZ{\bf Z}

\def\jb{{\bar j}}
\def\vf{{\varphi}}
\begin {document}
\centerline{\bf THE FRECKLED INSTANTONS}
\vskip 1cm
\centerline{A.~Losev $^{a}$, N.~Nekrasov\footnote{starting
in the fall 1999: Dept.
of Physics, Princeton University Princeton NJ 08544 USA}$^{,b}$,
S.~Shatashvili\footnote{On leave of absence from Steklov Mathematical
Institute, St. Petersburg, Russia}$^{,c}$} \vskip 1cm
\centerline{\em
$^{a,b}$ Institute for Theoretical and Experimental
Physics, Moscow,  Russia}
\centerline{\em $^{c}$ Physics Dept. Yale University,
New Haven CT 06540 USA}

\abstract{We study instanton-corrected renormalization group
flow in the two
dimensional sigma models and four dimensional gauge theory.
In two dimensions
we do that by replacing the non-linear supersymmetric
${\IC\IP}^{N-1}$ model
by the gauged linear sigma model which is in
the same universality class.  We
compare the moduli spaces of the instantons in
the non-linear model and that
of BPS field configurations in the linear model.
We reduce the problem of
matching of the parameters of the two systems to the intersection theory on
the compact moduli space of the latter model.
In four dimensions
we find that the
analogue of the linear sigma model is the gauge theory on the
non-commutative space-time.
Its BPS moduli space is the space of torsion free sheaves.
Both cases (2d and 4d) are unified by the notion of the
{\it freckled instantons}.  We also put an end to the discussion of the
nature of the superpotentials $W \sim {\s} {\rm log} {\s}$ in 2d and 4d and
discover the surprising disconnectnessness of the effective target space.
\vskip 1cm
\noindent
{\sl To appear in Yuri Golfand Memorial Volume}
\vskip 3cm
{\rm ITEP-TH-42/99}}
\vfill\eject
\section{Introduction}

The comparison between the two dimensional
sigma models and four dimensional gauge theories
is fruitful for both subjects\ci{polyakovi}.
Our paper continues
this line of the
research.

{\it Motivation.}
Our original motivation for the study is the interest in
the properties
of the renormalization group flow in the four dimensional
gauge theories.
This problem is both rich and has some chances
to be exactly
soluble in the context of the ${\CN}=2$ supersymmetric
gauge theories.

Consider for simplicity the case of the pure
               $SU(N)$ (twisted) supersymmetric
gauge theory.
The problem  is formulated
as follows. Let $\Phi$ be the complex adjoint scalar in
the vector multiplet.
For an irreducible representation $V$ of $SU(N)$
let ${\CO}_{V}  = {\Tr}_{V} \Phi$ be the
local operator in the gauge theory.

It commutes with the certain supercharge $Q$.
By acting on ${\CO}_{V}$
with other supercharges one can construct non-local gauge-
and $Q$-invariant
observables $\int_{C_{i}} {\CO}_{V}^{(i)}$, $i=0,1,2,3,4$,
where the
superscript $i$ denotes the degree of differential form,
which is
being integrated over a closed $i$-dimensional
submanifold $C_{i}$
of the space-time manifold.

The problem\ci{buckow} is now the following. The theory
has low-energy effective description in terms of the
${\CN}=2$ theory with
$r \equiv N-1$ abelian vector multiplets, whose scalars
$a^{i}, \quad
i = 1, \ldots, r$ are the special coordinates on the
moduli space of vacua
which is identified with the base of the family of the
hyperelliptic curves
${\CC}_{u}$:
\be
z + {{\Lambda^{2N}}\over z} = x^{N} + u_{1} x^{N-2}  +
\ldots + (-1)^{r+1} u_{r}
\label{crvs}
\ee
The parameters $u_{k}$ are identified with
the traces of $\Phi$ in the representations
$V_{k} = {\Lambda}^{k+1} {\IC}^{N}$:
\be
u_k = {\Tr}_{V_{k}}  \Phi
\label{trcs}
\ee

The problem is to find a representative for the operators
${\CO}_{V}$ and their descendants\ci{wittop}
$\int_{C_{i}} {\CO}_{V}^{(i)}$
in terms of $u_{k}$ and other data of the low-energy theory.
On the general grounds we expect to find a relation like:
\be
{\CO}_{V} \leadsto P_{V} ( u_{1}, \ldots, u_{r}; \Lambda)
\label{shata}
\ee
where $P_{V}(u_{1}, \ldots, u_{r}; \Lambda)$ is
the polynomial
whose value at $\Lambda = 0$ coincides with the classical
expression of the trace in the representation $V$ via the
traces in the fundamentals.
{\it This correspondence is  the four dimensional
generalization of the
well-known  quantum cohomology rings of two dimensional
supersymmetric sigma
models.}

{\it Scenario.} It was suggested in  \ci{NSVZ}
to study the renormalization group flow with
instantons present by integrating out all
fast fluctuating fields {\it a l\`a}
Wilson, including the small size instantons.
Now suppose we have a theory {\bf I} which has instantons
of all sizes, and a theory {\bf II} which has
both instantons of all sizes
and some other type of topological
defects which we shall call {\it freckles}
whose characteristic
size is bounded from above
by some parameter $\rho$. Then if we integrate out
all the fluctuations of the wavelengths smaller then
$\rho$ then the remaining
instantonic field configurations in both theories
{\bf I} and {\bf II}
become identical -- we say that the theories
are in the same universality class. If for some reason
the theory {\bf II} is simpler then the theory {\bf I} then
we can use the correlation functions in this
theory to compute the correlation functions in the
theory {\bf I}. One needs to express
the couplings $T_{*} ({\l})$ of the intermediate theory
at the scale ${\l} \gg \rho$
through the ultraviolet couplings
of the theories {\bf I} and {\bf II}:
\be
T_{*}({\l}) = f_{I}(T^{\bf I}_{*}; {\l}),
\qquad T_{*}({\l}) =
f_{II} (T^{\bf II}_{*}; {\l}), \qquad {\l} \gg {\rho}
\label{rg}
\ee
Upon exclusion of $T_{*}({\l})$ we arrive at the
relation:
\be
T^{\bf I}_{*}  = F ( T_{*}^{\bf II} ; {\l})
\label{rgi}
 \ee
Below we confine ourselves with the examples of
topologically twisted supersymmetric
sigma models in 2d and gauge theories in 4d.
In these cases (for $T_{*}$ being the topological
couplings, see below)
we expect the function $F( \cdot ; {\l})$ to
be independent
of ${\l}$ as long as ${\l}$ is greater then the
characteristic size $\rho$ of the freckles which
are specific for the model.

{\it Plan.} In the present paper we start to
carry out this program for
the supersymmetric
${\IC\IP}^{N-1}$ model in two dimensions
and then generalize to the four dimensional
gauge theory. The theories {\bf I} and {\bf II}
in these cases are respectively: non-linear
and gauged linear sigma model in two dimensions,
non-abelian gauge theory and (conjecturally)
the gauge theory on
the non-commutative space in four dimensions.
In the case of two dimensional theory we
find a subtle nature of the effective description of
the theory {\bf II}, namely we find that it
can be described as Landau-Ginzburg theory
on the disconnected space. Then we observe that
the similar phenomenon occurs in the compactification
of the four dimensional theory down to two dimensions
on a two-sphere.

\section{Two dimensional instantons with freckles}

Let us take for the theory $\bf I$ the two dimensional
supersymmetric non-linear ${\IC\IP}^{N-1}$ model.
It has
instantons represented by
the holomorphic maps $\varphi: \Sigma \to X$,
where
$\Sigma$ is the worldsheet
and $X \approx {\IC\IP}^{N-1}$
is the target space. The space ${\CM}$ of such instantons
is  disconnected and non-compact.
The former property  is due to the existence of
the maps of various degree $d \in {\pi} ({\Sigma}, X)
\approx
{\IZ}$ and it leads to the possibility of
adding a theta term to the action. The non-compactness
of ${\CM}$ is a serious albeit mostly computational
problem.

The correlation functions of certain observables
${\CO}_{i}$
in the theory
are the integrals over ${\CM}$ of certain differential
forms $\omega_i$. Was ${\CM}$ compact we could change
$\omega_i$ by the exact forms without affecting the integral
over ${\CM}$. This is no longer possible
for non-compact ${\CM}$. The situation is
not as innocent as one could think, since the
exterior derivative on ${\CM}$ is the
remnant of the (twisted) supersymmetry $Q$ of the model.
Usually one discards the $Q$-exact terms
in the correlations function when all the operators
are $Q$-closed.
This operation is dangerous when one has to regularize
integrals, precisely
because the virtual boundary contributions
may not vanish.

A way to compactify ${\CM}$ is provided by the
theory {\bf II} for which we take the gauged linear
sigma model with the gauge group $U(1)$ and
$N$ charged chiral multiplets\ci{witph}.

More generally, a theory $\bf II$ for the sigma model
with the target space $X \approx {\IC}^{N} // G$
is the gauged linear sigma model with $N$ chiral multiplets
which transform in a certain representation of the
gauge group $G$ and $//$ denotes the symplectic quotient
(the quotient of the space of zeroes of the $D$-terms
by the action of the gauge group $G$).

It turns out that the moduli space of the BPS field
configurations $\Psi$ (i.e. the solutions to the
equation $Q \Psi = 0$) in the gauged linear sigma
model is the compact space ${\overline\CM}$ which contains
the space ${\CM}$ as an open subspace.

\subsection{BPS moduli spaces in theory $\bf I$ vs. that of
\, \, $\bf II$}

In the case of $X \approx {\IC\IP}^{N-1}$
the space ${\overline\CM} = \amalg_{d} {\overline\CM}_{d}$
looks as follows:
$$
{\overline\CM}_{d} = \{ (P_0, \ldots, P_{N-1} ) \} / \sim
$$
where $P_{k}$ are the degree $d$ polynomials
( = holomorphic sections of the line bundle ${\CO}(d)$
over ${\Sigma}$ ) which are  not equal identically
zero altogether and $\sim$ is the equivalence relation:
$(P_{0}, \ldots, P_{N-1}) \sim ({\l} P_{0},
\ldots, {\l} P_{N-1})$, ${\l} \in {\IC}^{\star}$.
Thus ${\overline{\CM}}_{d} \approx {\IP}^{N(d+1)-1}$.
The simplicity of this space is misleading. The
catch
is that not every $N$-tuple of the polynomials
defines a map of ${\IP}^1$ to $X$. Only polynomials
without common divisors do that. Inside of
${\overline{\CM}}_{d}$ there is
a space ${\CM}_{d}$ of the polynomials
without the common zeroes. Unfortunately
in addition one finds a
stratum ${\CM}_{d-1} \times {\IP}^1$
which consist of
the polynomials  with one common zero, one also finds a
stratum ${\CM}_{d-2} \times {\IP}^2$
of polynomials with two common zeroes and so on,
all the way down to the stratum
$X \times {\IP}^{d}$ which consists of the polynomials
of the form $P_{k} (z) = a_{k} P(z)$
where $P(z)$ is an arbitrary degree $d$ polynomial  and
$$(a_{0} : a_{1} : \ldots : a_{N-1})$$ is a point
in $X$.  The common zeroes of
the polynomials $P_{k}$ are traditionally
called vortices. However we shall also call them
"freckles" for the reasons explained in the {\it scenario}.

In this way  we arrive at the
following stratification of
${\overline{\CM}}_{d}$:  \be {\overline{\CM}}_{d}
= {\CM}_{d} \cup {\IP}^1 \times {\CM}_{d-1} \cup \ldots \cup
{\IP}^{d} \times
{\CM}_{0}
\label{strat}
\ee
Following\ci{drinfeld} we refer to the points in
${\overline{\CM}}_{d}$ as to the {\it quasimaps}.
However having in mind the four dimensional
generalizations we suggest another name: the ``freckled
instanton''.

Notice that there are canonical maps
``gluing of the point-like instantons''
$$
v_{l}: \overline{\CM}_{d} \times {\IP}^{l} \to \overline
{\CM}_{d+l},
$$
which
actually add vortices-freckles to a quasimap:
\be
v_{l} \left( P_{0}(z), \ldots, P_{N-1} (z)\vert
x_1, \ldots, x_l \right) =
\left(P_{0} (z) Q(z), \ldots, P_{N-1} (z) Q(z)\right),
\label{glu}
\ee
$$
Q(z) = (z-x_1) (z - x_2) \ldots (z - x_l)
$$

\subsection{Great expectations: setup for the correlators}

A typical computation in the (twisted) supersymmetric
sigma model is the evaluation of the correlation
function of the observables
${\CO}_{\a}^{(0)}$ and $\int_{\Sigma} {\CO}^{(2)}_{\a}$
where ${\CO}_{\a}^{(0)}$ are the zero-observables
corresponding
to the cohomology classes of $X$, and
${\CO}^{(2)}_{\a}$ are their two-descendants.

To be specific let us take $X \approx {\IP}^{N-1}$.
The cohomology ring of $X$ is $N$ dimensional
and is spanned by $\o_r = \o^{r}, r = 0, \ldots, N-1$,
where $\o \in {\IH}^2(X)$ is the
K\"ahler form. The interesting observables
are the zero-observables
${\CO}_{r}^{(0)} \leftrightarrow \o_{r}$ and
the two-observables
$\int_{\Sigma} {\CO}^{(2)}_{r}, \quad  r >1$.
The two-observable of ${\o}_1 \equiv \o$
measures the degree of the
holomorphic map:
$$
\langle \ldots \int_{\Sigma} {\CO}^{(2)}_{1} \rangle_{d} = d
\langle \ldots \rangle_{d}, \quad d = \int_{\Sigma}
{\varphi}^{*} \o
$$
which is constant in the given instanton sector.

The correlation function
\be
\langle \CO_{{\a}_{1}}^{(0)}(x_1) \ldots
\CO_{\a_{k}}^{(0)}(x_k) \int_{\Sigma} \CO^{(2)}_{\b_{1}}
\ldots \int_{\Sigma}   \CO^{(2)}_{\b_{l}} \rangle
\label{cornlsm}
\ee
computes {\it the number of the
rigid
holomorphic maps of $\Sigma$ to $X$ with the following
conditions:
the fixed points $x_1, \ldots, x_k$ on $\Sigma$
are mapped to the submanifolds  $C^{\a_{1}},
\ldots, C^{{\a}_{k}} \subset X$ which represent the cycles
which are Poincare dual to the cohomology classes
$\o_{\a_{1}}, \ldots, {\o}_{\a_{k}}$ of $X$
while some points $y_1, \ldots, y_l$
whose position is not specified are mapped
to the submanifolds
$C^{\b_{1}}, \ldots C^{\b_{l}} \subset X$
whose homology classes are dual to
$\o_{\b_{1}}, \ldots, \o_{\b_{l}}$}.

The answer is independent of the specific choice
of $C^{\a}, C^{\b}$ as long as everything remains
in generic position\footnote{
This interpretation of the
correlation function holds when one chooses
the specific representatives for the cohomology
classes $\o_r$ of $X$, namely delta functions
supported at $C^{\a}, C^{\b}$. If one smoothes
out the delta functions then at some point
the supports of the smeared delta functions
overlap and one should use more intricate arguments
to show that the correlation function computes the same
thing.}.

In order to use the compactification ${\overline{\CM}}_{d}$
one has to formulate the computation of the correlation
function \re{cornlsm} in terms of the linear gauged
sigma model, more specifically, in terms
of the $N$-tuples of the polynomials of degree $d$.

It is not hard to do that.
The cycle in ${\IP}^{N-1}$ which is Poincare
dual to ${\o}_{r_{\a}}$ can be represented
by a plane $C^{\a}$
which is the space of solutions
to a system of linear equations
${\ell}_{i}^{\a} = 0$, $i=1, \ldots, r_{\a}$ and each
${\ell}_{i}^{\a}$   is a section of ${\CO}(1)$, i.e.
the linear function in the homogeneous coordinates
$w_{0}, \ldots, w_{N-1}$ on $X$:
$$
{\ell}_{i}^{\a} (w_{0}, \ldots,
w_{N-1}) = \sum_{\kappa =0}^{N-1} L_{i,\kappa}^{\a}
w_{\kappa}, \qquad
L_{i,\kappa}^{\a} \in {\IC}
$$
Now the correlation function in the non-linear
sigma model
computes the number of such
$N$-tuples
$P_{0}, \ldots, P_{N-1}$ modulo the common multiple
${\l} \in {\IC}^{\star}$
and the points $y_1, \ldots, y_l \in {\Sigma}$
 which obey the following equations:

\be
{\ell}_{i}^{{\a}_{p}} \left(
P_{0}(x_{p}), \ldots, P_{N-1}(x_{p})\right) = 0,
\quad i = 1, \ldots, r_{{\a}_{p}}, \qquad p = 1, \ldots k
\label{zrobs}
\ee
\be
{\ell}_{i}^{{\b}_{q}} \left(
P_{0}(y_{q}), \ldots, P_{N-1}(y_{q})\right) = 0
\quad i = 1, \ldots, r_{{\b}_{q}}, \qquad q = 1, \ldots l
\label{twoobs}
\ee

{\it If we relax the conditions on the polynomials $P_{k}$,
i.e. allow them to have common divisors we
replace the computation in the non-linear sigma
model by that in the gauged linear sigma model.}

The drastic difference between the spaces
${\CM}_{d}$ and ${\overline{\CM}_{d}}$ shows up in
the following: if the polynomials
$P_{0}, \ldots, P_{N-1}$ have a common
zero $x_{*}$ then the equations
\re{twoobs} have solution e.g.
$y_{a} = y_{b}  = y_{c} = x_{*}$
for some $a \neq b \neq c$.  These solutions have nothing
to do with the properties of the actual
holomorphic map defined by the polynomials $P_{k}$
with the common factor divided out. In other words
they do not contribute to the solution of the
enumerative problem posed by the non-linear
model. Nevertheless they contribute to the correlation
function of the linear model.

{\it The accurate subtraction of these extra contributions
is the heart of the relation \re{rg}
between the non-linear and the linear sigma models. The most
notorious example of this renormalization is
known under the name of the mirror map in the case
of sigma models with Calabi-Yau threefolds as the
target space\ci{witph,giv,plmor}.}

In case of the
manifold $X \approx {\IP}^{N-1}$ (more generally
if $X$ is a $D$-dimensional
Fano variety with $\int_{X} \o^{D-1} \wedge c_{1}(X) > D$)
the correlations functions of the zero-observables
${\CO}^{(0)}_{\a}$ can be rather easily computed
by replacing the space ${\CM}_{d}$
of the holomorphic maps by the space ${\overline{\CM}}_{d}$.
The proof goes as follows:
Suppose that the ghost numbers ${\Delta}_{{\a}_i}$
of the observables
${\CO}_{{\a}_{1}}^{(0)}, \ldots, {\CO}_{{\a}_{k}}^{(0)}$
(which coincides with the degree of the cohomology classes
of $X$ which they represent) saturate the
ghost number anomaly in the instanton sector $d$, i.e.
$$
\sum_{l} \Delta_{a_{l}}  = N(d+1) - 1
$$
Since the positions of the zero-observables $x_{1},
\ldots, x_{k} \in {\Sigma}$ are distinct the
vortices can contribute to the correlation function
$\langle
{\CO}_{{\a}_{1}}^{(0)}(x_1) \ldots
{\CO}_{{\a}_{k}}^{(0)}(x_k) \rangle$
only if their location coincides
with some  of the points $x_1, \ldots, x_k$.
The vortices with the fixed
location $x_{i_1}, \ldots, x_{i_p}$ on $\Sigma$ form
a submanifold
$v_{p} ( {\CM}_{d-p} \times \{ x_{i_1}, \ldots, x_{i_p} \} )
\subset {\overline{\CM}_{d}}$ of codimension $N p$.
Each class in $H^{*}(X)$
has degree which is less then $N$ (in complex units).
The complement to
the vortices in $\Sigma$ is mapped to $X$ with the degree
$d-p$
therefore the
rest of the observables must be saturated by
the zero modes which are present
in the holomorphic maps of the degree $d-p$,
i.e. by $N ( d- p+1)-1$ zero
modes.  But their total  number is
$$ \sum_{l \neq i_{1}, \ldots, i_{p} }
{\Delta}_{l} \geq N(d+1) - 1 - p (N-1) =
$$
$$ =  N ( d-p+1) -1 +  p > N ( d- p + 1) -
1 = {\rm dim}{\CM}_{d-p}.
$$
Hence the dimensions do not match and the
vortices do not contribute.

The counting changes as soon as we start computing
the correlation
functions of the two-observables.
In that case the point of insertion of the two-observable
is not fixed and may hit other observables at the same
time with the vortex. If the total  ghost number of the
collided observables is greater then $N-1$ then
in general there is  a  contribution of the vortex to the
correlation function. {\it It is this phenomenon which
induces the
contact terms\footnote{It is well-known
that the contact terms correspond to the change of
couplings\ci{polyakov}.} which enter the relation
\re{rgi} and provides its microscopic explanation}.

\subsection{Computations in theory $\bf II$:
geometric story}
Let us compute the generating
function of the correlators of the zero- and two-observables
in the gauged linear sigma model.
The arguments of the
equations \re{zrobs}, \re{twoobs} are the polynomials
$P_k$ and the points $y_{q}$. The equations
themselves are
the conditions of the vanishing
of certain  sections of the vector bundles
\footnote{
The bundles and their r\^ole
are the following: the zero-observables
${\CO}^{(0)}_{\a_{p}}$ represented by
\re{zrobs}  is the top degree
Chern class $c_{top}({\CE}_{r_{\a_{p}}}^{0})$
of the bundle ${\CE}_{k}^{0} = {\pi}^{*} {\CO}(1) \otimes
{\IC}^{k}$, while
the integrated two-observable
$\int_{\Sigma} {\CO}^{(2)}_{r_{\b_{q}}}$
is the integral over the $q$-th copy $\Sigma$ in
$\Sigma^{l}$ of the top degree Chern class of the
bundle ${\CE}^{2}_{k,q} =
{\pi}^{*} {\CO}(1) \otimes p_{q}^{*} {\CO}(d) \otimes
{\IC}^{k}$. Here
${\pi}, p_q$ are  the projections
${\overline\CM}_{d} \times \Sigma^{l} \to
{\overline{\CM}}_{d},$
and to the $q$-th copy of $\Sigma$ respectively}
${\CE}^{0}_{r_{\a_{p}}}, {\CE}^{2}_{r_{\b_{q}}}$ over
${\overline{\CM}}_{d} \times
\overbrace{\Sigma \times \ldots \times
\Sigma}^{l \, \rm times}$.
Clearly, inside of a correlator in the instanton
sector with the instanton number $d$ the following relation
holds: $$
\int_{\Sigma} {\CO}^{(2)}_{k} = k \int_{\Sigma}
{\CO}^{(2)}_{1} {\CO}_{k-1}^{(0)} = k {\CO}_{k-1}^{(0)}
\int_{\Sigma} {\CO}_{1}^{(2)} = kd \cdot {\CO}_{k-1}^{(0)} $$
(this statement follows from
the K\"unneth
formula \ci{issues}, cf. \ci{witdgt}).
After these preparations we are now in the position
to write down the answer:
\be
{F}^{\rm lin}_{0} (t, T) : =
\sum \prod_{k=0}^{N-1} \frac{T_{k}^{n_{k}} t_{k}^{m_{k}}}{
n_{k}! m_{k}!}
\langle \prod_{k= 0}^{N-1} \overbrace{\CO_{k}^{(0)}}^{m_{k}}
\overbrace{
\int_{\Sigma}  {\CO}^{(2)}_{k} }^{n_{k}} \rangle =
\label{genfn}
\ee
$$
= \sum_{d,l} \int_{{\overline{\CM}}_{d} \times {\Sigma}^{l}}
\exp \left( \sum_{k} t_{k} c_{top} ({\CE}_{k}^{0})
+ \sum_{q} T_{r_{q}} c_{top} ({\CE}_{r_{q}, q}^{2}) \right) =
$$
$$
= \sum_{d,l} \oint \frac{d\s}{\s^{N(d+1)}}
\prod_{q=1}^{l} \frac{d \o_{q}}{\o_{q}^2}
\exp \left( \sum_{k} t_{k}  \s^{k} + \sum_{q, \{ r_{q} \}}
T_{k_{q}} \left(
\s + d \cdot \omega_{q} \right)^{r_{q}}   \right) =
$$
$$
= \sum_{d} \oint \frac{d\s}{\s^{N(d+1)}}
\exp\left( \sum_{k} t_{k} \s^k + kd \cdot T_{k} \s^{k-1}
\right) =
$$
\be
= \oint
\frac{d\s \, \exp \left( \sum_k t_k \s^k\right)}{{\s}^{N} -
\exp \left( \sum_r r T_r {\s}^{r-1} \right)}
\label{cntr}
\ee
This representation of the generating function of
the correlation functions in the linear gauged sigma
model is rather suggestive yet shows the crucial
difference between the non-linear and the linear
gauged models. The difference shows up as
the asymmetry between the zero- and two-observables and
the absence of the constant (in $T_k$) metrics
 on the space of    zero-observables \ci{km,wdvv}.
The suggestive part of
the formula \re{cntr} is in
its contour integral form,
which looks very much like the
correlation function in the topological
Landau-Ginzburg theory\ci{vafalg}
(with the contact terms\ci{asl}
equal to
zero).
The latter must have the following structure:
\be F^{\rm LG}_{0} (t,T) =
\oint \frac{\mu^2(X) d X^1 \wedge \ldots \wedge d X^D  \,
\exp \sum_k t_k \Phi_k (X)}{\p_1 W_{T}(X) \ldots
\p_D W_T (X)} \label{LG} \ee where $X^i$
are the holomorphic coordinates on the target space $\CX$
of the LG sigma
model,
$dW_T(X) = dW_0(X) + \sum_k T_k d\Phi_k$ is the holomorphic
one-differential (the derivative of
the superpotential $W_T$), $\Phi_k$ are
the representatives of the local ring of the superpotential
and $\mu(X) dX^1
\wedge \ldots \wedge dX^D$ is the holomorphic top degree
form on $\CX$
\ci{vafalg,asl,witmir,krichever}.  The formula \re{cntr}
 is not  exactly of
the form \re{LG} but it is easy to map it into this form:
replace the
summation over $d$ by
the integral using the Poisson resummation trick:
$$
\sum_{d \geq 0} A_{d} = \sum_{n \in {\IZ}} \int_{0}^{\infty}
e^{2\pi i x n}
A_{x}  dx
$$
assuming that the function $A_{d}$ defined on ${\IZ}_{+}$
can be
extended to the whole of ${\IR}_{+}$ as a continious function.
In order to
apply this trick to the formula above\re{cntr}
 we rewrite the contour
integral over $\s$ as the integral over
$\varphi$ from $0$ to $2\pi$ with $\s
= e^{i\varphi}$.
Now we define $\s^{x}$ as $e^{ix\varphi}$.
Doing these
manipulations we
arrive at:
$$
F^{\rm lin}_{0}(t, T) =
$$ $$
\sum_{n \in \IZ} \int_{0}^{2\pi} d\varphi
\int_{0}^{\infty}  dx e^{i\varphi
(1 - (x+1) N)+ 2\pi i n x}
\exp
\left(
\sum_{k} t_{k} e^{ik\varphi}
 + k x T_{k} e^{i(k-1)\varphi} \right)
=
$$
\be
= \int_{-\infty}^{\infty} d\varphi \sum_{n =0}^{N-1}
\frac{d\varphi \, \exp \left( i \varphi ( 2 - N) +
\sum_k t_k e^{ik\varphi} \right)}{\left(
2\pi i n - i N \varphi \right)
e^{i\varphi} + \sum_k k T_k e^{ik \varphi} + i0} =
\label{presu}
\ee
\be
\oint_{\CX} \frac{\mu^2 (\varphi)
d\varphi e^{\sum_k t_k \Phi_k
(\varphi)}}{\p_{\varphi} W_{T}}
\label{LGCP}
\ee
where $\CX = \overbrace{\IC
\amalg \ldots \amalg \IC}^{N \, \rm times}$
is the disjoint union of $N$ copies of the complex
line, $\mu d\varphi = e^{i ( 1- N/2) \varphi} d\varphi$,
$\varphi \in \IC$,
the observables $\Phi_k = e^{ik\varphi}$,
and the superpotential on the
$n$'th copy of $\IC$ is equal to:
\be
W_{T} (\varphi) = - \left( N \varphi -
2\pi n \right) e^{i\varphi}        +
\sum_{k} {\tilde T}_{k} e^{ik \varphi},
\qquad i {\tilde T}_{k} = T_{k} + N \delta_{k,1}
\label{LGsu}
\ee
The contour integral in \re{LGCP} is taken around
infinity at each connected component of ${\CX}$.
As usual the expression \re{LGCP} must be understood
perturbatively in $T_k$ for $k > 1$.

The requirement that $W_T$ must be linear in $T_k$
constrains it but does not fixes it uniquely.
We could interpret \re{presu} in a
slightly different
way, by taking $\mu = e^{i(-N/2)\vf}$
and
${\tilde W}_{T} = - \frac{1}{2}
\left( N \vf - 2\pi n \right)^2  + T_{1} \vf +
\sum_{k=2}^{N-1} \frac{k}{k-1} T_k e^{i(k-1)\vf}$.

It is not possible to make a fair choice between
all these options without further physical insight.
We turn to it in the next subsection.

\subsection{Physical derivation}

Despite the exotic nature of the effective
target space and the superpotential
which we derived in the previous
section using topological techniques applied
to the moduli space of the BPS field configurations
in the gauged linear sigma model it makes perfect sense!
In fact, we shall now derive the same result using
the effective field theory techniques.

Recall the field content of the
two dimensional gauged linear $\CN=2$
supersymmetric sigma model. The chiral multiplets
$\Phi^i =
\left( \phi^i, \psi^{i}_{\pm}, F^i \right)$
take values in some complex
vector space $W$ with the constant K\"ahler form
$\o$.
The vector multiplets
$V= \left( A_{\m} , \s, \bar\s, \l_{\pm},
\bar\l_{\pm}, D \right)$
take values in the Lie algebra of the
group $G$ which acts in $W$ preserving K\"ahler structure.
The superfield $\Sigma$ which
contains the field strength ${\CF} = dA + \frac{1}{2} [A,A]$
is the twisted
chiral superfield with the quantized highest
component $F$ (for abelian $G$).

The bosonic part of the Lagrangian is
(in the absence of the tree level superpotential):
\be
\int \Vert D_{\m} \phi^i \Vert^2 +
{1\over 2e^2} F_{\m\n}^2 +
{e^2 \over 2} \Vert \mu \Vert^2 +
\Vert [ \s, \bar\s ] \Vert^2
\label{bos}
\ee
where $\mu \sim T^{a}_{i\jb} \phi^{i} \bar\phi^{\jb}$ is
 the
moment map for the $G$ action in $W$.
If the gauge group
$G$ contains $U(1)$ factors then we may
deform the model by adding a constant per each $U(1)$
to $\mu$ (Fayet-Illiopoulos terms)
thereby promoting $\mu$ to
\be
\mu =   T^{a}_{i\jb} \phi^{i} \bar\phi^{\jb} - r_i {\bf 1}_i
\label{flmu}
\ee
Also,
for each $U(1)$ factor there is a
$\t$ term: $\t_i \int F_i$.
Altogether we get a complex parameter
$t_{i} = ir_i + {{\t}_i \over 2\pi}$
 per each $U(1)$.
Correlation functions of chiral observables
are {\it holomorphic}
in $t_{i}$ -- an important constraint.

Finally,
$e^2$ in \re{bos} is the gauge coupling (for
several $U(1)$ factors one may
have different couplings $e_i^2$).
The standard lore says that
in the infrared $e_i^2 \to \infty$ and the gauged
linear model
looks more and more like
the non-linear sigma model with the target
$$
X= W//G = \mu^{-1}(0)/G
$$
For $G = U(1)$, $W = {\IC}^{N}$ being the $N$ times the
standard charge one representation of $G$ the space
$ X$ is isomorphic to ${\IC\IP}^{N-1}$ for $r > 0$,
$X$ is a point for $r = 0$ and $X$ is empty for $r < 0$.

The field configurations which preserve some of the
supersymmetry obey the equations \ci{witph}:
\be
\left( \pb + {\bar A}  \right) \phi^i = 0, \qquad
F = - e^2 \mu
\label{susy}
\ee
It is possible to show that for the spherical
worldsheet the space of solutions to these equations
in a given instanton sector ($\int F = 2\pi i d$)
coincides with the space $\overline{\CM}_{d}$ of quasimaps\ci{stable}.
The quasimaps with vortices correspond to the perfectly
smooth solutions of \re{susy} which are the generalizations
of the Abrikosov-Nielsen-Olesen vortices\ci{ano}.
They have the
characteristic
size of the order $\frac{1}{e \sqrt{r}}$ and therefore
for $\rho > \frac{1}{e \sqrt{r}}$ are integrated out
in \re{rg}. The fact that the vortices have the finite size
in the microscopic theory $\bf II$ is another
justification for the
name ``freckle'' which we suggest.

To derive the relations \re{LGCP}, \re{LGsu} within the
physical theory one integrates out the $N$ chiral
multiplets in the background of  slowly varying
fields of the vector multiplet. One induces
the well-known
twisted superpotential\ci{dadda} (in our notations)
\be
{\widetilde W}_{\rm eff} ({\s}) = N \frac{\s}{2\pi}
{\rm log} \frac{\s}{\tilde \Lambda} \label{twsu} \ee
with $\tilde \Lambda = e^{- 2\pi
it -1}$.  The complex scalar $\s$ enters the twisted chiral superfield
$\Sigma$ which has a constrained $F$ component - it comes from the gauge
field strength hence must be quantized:  \be \int F \in 2\pi i \IZ
\label{qua} \ee
If this was not the case the superpotential \re{twsu} would
not make any sense due to the branching of the logarithm\ci{witph}.
The
bosonic part of
the Lagrangian which contains the effective superpotential
\re{twsu} has the form:
\be \int {\p}_{\s} {\widetilde W}_{\rm eff}  \left(
D + i F \right) +
{\overline{{\p}_{\s}\widetilde W}}_{\rm eff} \left( D - i
F \right)
\label{bosi}
\ee
To draw some physical
conclusions from the shape of the effective
superpotential we need
 to map
the constrained superfield $\Sigma$ to the unconstrained
(twisted) chiral multiplet. Otherwise we cannot simply
integrate out $F$ in \re{bosi}
to get an effective potential.

Instead we perform
the following duality transformation.
First of all we assume the worldsheet to have a
spherical topology (it is not necessary
but simplifies the discussion a bit).
Let us choose an arbitrary point $p$ on the worldsheet.
The field ${\s}$ taking values in ${\IC}^{\star}$
(the point $\s = 0 $ is the singular point as there
the matter fields become massless and
this singularity is
avoided dynamically as we shall see in a minute)
is the same thing as the field $\vf$ taking values
in $\IC$, $\s = \exp i \vf$ with the only
condition:
\be
0 \leq {\rm Re} \, \vf (p ) < 2\pi \label{gge}
\ee
This condition is not holomorphic but this will not
bother us, as we will  get rid of it very soon.
The change of variables from $\s$ to $\vf$ with the
constraint \re{gge} maps all the   fields in the supermultipet
$\Sigma$ to those in the supermultiplet $\Phi$.
Now we can relax the condition \re{qua} at the expense
of adding to \re{bosi} a term
\be
2\pi i n \int F
\label{pois}
\ee
where $n \in \IZ$ is the integer-valued ``field''
which must be summed over in the path integral.
The term \re{pois} is equivalent to the shift of the
twisted superpotential by the term, linear in $\s$:
$$
{\widetilde W} \to {\widetilde W} +
2\pi i n \s  = i  e^{i\vf} \left( N \vf - N \tilde t -
2\pi  n
\right)
$$
Now,
the summation over all $n \in \IZ$
with the path integral over $\vf$ obeying \re{gge} is equivalent to the
summation over $n \in {\IZ}_{N}$ with the path integral over {\it all} $\vf$
without any constraints, as promised. For
the shift of $\vf$ by $2\pi$
is equivalent to the shift of $n $ by $N$.
In this way we arrive at the formula \re{LGsu} with
$T_1 = N t$, $T_k = 0 $, $k> 1$.

To get the full set of ``times'' $T_k$ we must start with the
linear sigma model deformed by all two-observables.
It is easy to see that this deformation is equivalent
to the shift of the superpotential by the terms
$\sim T_k {\s}^k$, $k > 1$.

\subsection{More examples of disconnected targets}

In this subsection
we consider more examples of effective theories
with disconnected target spaces. This story slightly
takes us off the route so the reader who is interested
in further details concerning the relations
between the theories $\bf I$ and $\bf II$ may skip it
and proceed directly to the next section.

{\it Compactification of ${\CN}=2$ gauge theory.}
Consider four dimensional ${\CN}=2$
supersymmetric Yang-Mills
theory with the gauge group $SU(N)$.
Let us keep the four dimensional coupling finite so that
we allow the point-like instantons to contribute to the effective
action. Then the effective two-dimensional theory
is given simply by the Kaluza-Klein reduction
of the effective four dimensional low-energy action.

Recall that the bosonic part of the latter is given by
the following expression:
\be
S_{bos} = \int {\tau}_{ij} F_{i}^{-} \wedge F_{j}^{-} -
{\bar \tau}_{ij}
F_{i}^{+} \wedge F_{j}^{+} +
{\rm Im}\tau_{ij} da^{i} \wedge \star d{\bar a}^{j}
\label{eff}
\ee
The couplings $\tau_{ij}$, $i,j = 1, \ldots, N-1$
are determined with the help of the
family of hyperelliptic curves \re{crvs}.

When we consider all possible configurations
of the gauge field
we must include those which have non-trivial fluxes
through the sphere ${\bf S}^2$. Naturally we get sectors labelled
by $\vec n = ( n_{1}, \ldots, n_{N-1})$,
where
\be
\int_{{\bf S}^2} F_{i} = 2\pi i n_{i}, \quad n_{i} \in {\IZ}
\label{flx}
\ee
Now, the resulting two dimensional theory still contains abelian
 gauge fields. As it is well-known the gauge fields are non-dynamical
in two dimensions. It is perhaps
less known that this non-dynamical
nature of the gauge fields can be
conveniently summarized
by saying that instead of the path integral over
the gauge equivalence classes of the gauge fields
one can simply take the path integral over two-forms
$F_{i}$ with the only constraint that the integral
of $F_{i}$ over the (compact) space-time is
quantized\ci{witdgt,thompson,anton}:
$$ \int_{\rm space-time} F_{i} =
2\pi i {\tilde m}_{i}, \quad {\tilde m}_{i} \in {\IZ} $$ This description
automatically takes care of the gauge fixing and the Faddeev-Popov
determinants. As we did in the previous analysis
it is convenient
to enforce the latter constraint  by introducing
another integer-valued field
$\vec m = (m^{1}, \ldots, m^{N-1}), \, m_k \in {\IZ}$
to be summed over,
relaxing the condition on $F_{i}$
by allowing it to be any two-form,
at the same time  adding to the action an
extra term
\be
S \to S + \int_{\rm space-time} m^{i} F_{i}
\label{dual}
\ee

Let us compactify this theory
on a two-sphere ${\bf S}^{2}$. Since the sphere has no
covariantly constant spinors
the supersymmetry will
be broken. In order to avoid that
we consider the (partially) twisted
theory, i.e. add certain curvature couplings
to the Lagrangian. The simplest approach starts
with the theory which is
topologically twisted in four dimensions
\ci{wittop,lauriz}. The field
content of the theory is:
a vector $A$, a one-form fermion $\psi$,
a self-dual two-form fermion
$\chi$, a scalar fermion $\eta$, an auxiliary
self-dual bosonic
two-form $H$, a complex bosonic scalar $a$.

Upon Kaluza-Klein reduction (which amounts
to keeping the internal
parts of the fields harmonic) we will get:
a vector $A$, a one-form fermion $\psi$, a two-form fermion
$\chi$, a scalar fermion
$\eta$, an auxiliary two-form bosonic field  $H$, a complex
bosonic scalar $a$.
Upon the trick with the introduction of the auxiliary labels
$\vec m$ described in the previous section the
vector $A$ is replaced by another auxiliary bosonic
two-form $F$. All fields have in addition a label $i$ which runs
from $1$ to $N-1$.
Let us look at the supersymmetry transformations: the scalar
supercharge $Q$ acts as follows:
\begin{eqnarray}
& Q F_{i} = d \psi_{i}, \quad Q \psi_{i} = d a_{i}, \quad Q a_{i} = 0 \\
& Q \chi_{i} = H_{i}, \quad Q H_{i} = 0, \quad \\
& Q \bar a^{i} = \eta^{i}, \quad Q \eta^{i} = 0
\label{susyi}
\end{eqnarray}
One can easily recognize here the $Q$-operator of the
topological type $\bf B$ sigma model.
Usually the fermion $\chi_{i}$ is denoted as $\theta_{i}$, while the
auxiliary field $H_{i}$ goes under the name ${\bar F}_{i}$.

{\it What is the target space?}
The immediate answer would be: ``the space of $a_{i}$'s'', since these
are the target space coordinates in the
supersymmetry transformation laws. But we should remember
that on the route to the sigma model description we were
introducing some extra labels $\vec n, \vec m$ when we dealt
with fluxes of the gauge field through the internal sphere as well as
the two dimensional space-time itself.

The second guess would be that we should take as many copies
of the space of $a_{i}$'s as there are labels $(\vec n, \vec m)$.
The truth sits in between.
Recall that $a_{i}$'s
are only {\it local}
special  coordinates on the moduli space $\CV$
of vacua of the gauge theory.
As one goes around some non-contractible loops in $\CV$
one comes back with another
set of $a_{i}$'s, transformed by a monodromy group.
This very group
also transforms the vectors $(\vec n, \vec m)$.

After all these symmetries are taken into account we get
the following statement:

{\it The target space $V$ of the effective type }{\bf B} {\it sigma model
is the space of pairs:} $({\CC}_{u}, \gamma)$, {\it where }
${\CC}_{u}$
{\it is one of the curves} \re{crvs} {\it and } $\gamma \in
{\IH}_{1}({\CC}_{u}, {\IZ})$.

The unusual property of this target space is its
disconnectness: the cycles $\gamma_1, \gamma_2$ which are
not in the same orbit of the monodromy group belong
to the different connected components. The component
where $\gamma = 0$ is isomorphic to the
moduli space $\CV$ of the curves \re{crvs}, while
the other components are certain covers of $\CV$:
$$
V \approx {\CV} \amalg {\widetilde{\CV}} \amalg \ldots
$$
For example, for $N=2$, the
component with $\gamma = 0$ is isomorphic to the
complex line with the coordinate $u$, parameterizing
the curves:
$$
y^2 = (x^2 + u)^2 - 4\Lambda^4
$$
while the other components are isomorphic to the
strip $0 \leq {\rm Re}\,\tau \leq 4, {\rm Im}\,\tau > 0$.
This space shows up in certain computations in Donaldson
theory \ci{moorewit,issues}.

{\it Superpotential.}
The curves \re{crvs} are embedded by definition into
a hyper-k\"ahler manifold $T^{*}{\IP}^1$, which has
a holomorphic two-form $\omega$:
\be
\omega = \frac{1}{2\pi i} dx \wedge \frac{dz}{z}
\label{smpl}
\ee

Our sigma model has a (twisted) superpotential $W$. In general,
the superpotential needs not to be well-defined. Only
the derivative $dW$ must be a well-defined
holomorphic one-form on the target space.

In our case we can write a simple formula for
the derivative $dW$:
\be
dW = \oint_{\gamma} \omega
\label{super}
\ee
To prove the last assertion let us introduce a few more
standard notations. Let $\lambda = \frac{1}{2\pi i} x \frac{dz}{z}$
be the meromorphic one-differential, $d\lambda = \omega$.
Its residues on the curve ${\CC}_{u}$
vanish for any $u$. Hence we can integrate \re{super}
to get:
\be
W = \oint_{\gamma} \lambda = n_{i} a^{i} + m^{i}a_{D, i}
\label{inte}
\ee
where we expanded $\gamma$ in some (local) symplectic basis $(A_{i}, B^{i})$
in ${\IH}_{1}({\CC}_{u}, {\IZ})$ and
$$
a^{i} = \oint_{A_{i}} \lambda, \quad a_{D,i} = \oint_{B^{i}} \lambda
$$
Now let us compute the potential which follows from the
superpotential \re{inte}:
\be \left(  n_{i} + \tau_{ik} m^{k} \right)
\left( {\rm Im}\tau^{-1}\right)^{i\jb}
\left( n_{j} + \bar\tau_{jl} m^{l} \right)
 \label{actn} \ee
This is  precisely the potential which one would
get from the
gauge theory action
upon compactifying on the small sphere and
performing the two dimensional duality transformation.
Of course in this very setup one also has Kaluza-Klein
modes along the two-sphere which have the same order of energy
as the fluctuations in the potential \re{actn}. One can
separate the two scales by adding the two-observable
$t {\CO}^{(2)}_{u}$
whose support extends over all of the two dimensional
space-time. This addition changes the superpotential
by    $t u$ and introduces another scale into the problem.
More details on this issue can be found in \ci{issues}.

{\it Remarks.} The effect of
the unfolding of the moduli space
of twisted chiral multiplets is very general. In fact,
one can
use the similar method of unfolding in four dimensional theories
\ci{kovnershif,kss} with effective superpotential
of Veneziano-Yankielowicz type\ci{venyan}. To the best of
our knowledge
the phenomenon of creating
of disconnected components
in the effective target space
was not observed before
neither in two- nor
in four-dimensional models.
This phenomenon makes the study of
solitons connecting different vacua rather
intricate. For the
solitons are represented by the paths
connecting the vacua sitting on
different components of the target space.
It means that the trajectory
connecting the vacua should break at some point.
The only way this breaking may be avoided is to
assume that
the metric on the effective target space is such that
the components are actually glued together
at infinity.

If this is not the case then the actual
location of the breaking point on each sheet
might
be undetermined by means of the low
energy theory alone. This is similar
to the fact that the
mass of the magnetic monopole cannot be computed from the
Maxwell theory alone.
We shall elaborate on this and related issues
elsewhere \ci{progr}.

Notice that if the model is embedded into
the string/M-theory, e.g. via brane realization
then one gets a geometrical representation of
the solitons\ci{horihanany}. Also note that the
superpotential of the SW form (but without unfolded
disconnected target space) was computed in \ci{lerche}
in the geometrical engineering approach to the
compactifications on the Calabi-Yau
fourfolds. We plan to return to this question elsewhere.

\subsection{Comparison to the results of
the XIX century}

Despite the success of the previous two subsections
we should warn the reader that their results where
merely the
groundwork for the solution of the real problem -
the effective theory description of the
non-linear sigma model.
To see that the renormalization\re{rg},\re{rgi}
is non-trivial
let us perform several numerical checks.

To be specific let us take $N=3$, i.e. ${\IC\IP}^2$
model. The genus zero
correlation functions in the topologically
twisted  theory compute the numbers of
rational planar  curves  which pass through
a given number of points in ${\IP}^2$ and intersect
a given number of lines in generic position.
In particular:
\be
\langle \CO^{(0)}_{1} (0) \CO^{(0)}_{1}(1)
\CO^{(0)}_{1} (\infty) \int \CO^{(2)}_{2} \int
\CO^{(2)}_{2} \rangle^{\rm non-lin} = 1
\label{ooott}
\ee
is the number of degree one curves (lines) which
pass through two distinct points and intersect
three lines in generic position. From the elementary
school we know that this number is equal to one.
On the other hand, it is easy to see that
\be
\langle \CO^{(0)}_{1} (0) \CO^{(0)}_{1}(1)
\CO^{(0)}_{1} (\infty) \int \CO^{(2)}_{2} \int
\CO^{(2)}_{2} \rangle^{\rm lin} = 2^2 = 4
\label{looott}
\ee
This is the first in the series of numbers
predicted by the gauged linear  sigma model
which  differ from those computed by
the last century algebraic geometers.

Of course, the difference happens precisely when we
start looking at the  correlation functions of
two-observables, as we promised in the beginning
of the section. We also said that the difference
is due to the contribution of the freckles\footnote{We should
mention that the similar effects
were observed more
then a hundred years before\ci{cayley} the discovery
of supersymmetry\ci{golfand}}.
Let us now
see explicitly how this happens.

The space ${\overline\CM}_{1}$ contains the subspace
${\IP}^1 \times {\IP}^2$ of vortices, where ${\IP}^1$
parameterizes the location $x_{*}$  of the vortex
and ${\IP}^2$ is the set of images ${\bf r}$
in the target space of
the complement to the vortex. The equations
\re{twoobs},\re{zrobs} are obeyed both by the proper
map of degree one and by the vortex configurations
with $x_{*} = y_1 = y_2 = x_{p}$ where $x_{p}$
assumes one of the three available values: $0,1,\infty$.
The two remaining equations state that the image
$\bf r$ must belong to two lines
${\ell}_{p^{\prime}}, p^{\prime} \neq p$
in ${\IP}^2$, i.e. to their intersection point.
Altogether we found three vortex configurations,
each contributes one to the correlation function, hence:
\be
4^{\rm lin} = 1^{\rm non-lin} + 3^{\rm vortex}
\label{fp}
\ee
The similar (but much more tricky) counting
works for higher instanton charges. The effect of
subtraction of the vortex contributions is to replace the
effective superpotential
\re{LGsu} by the infinite series
\be
\left( N {\vf} - 2\pi n - T_1\right) e^{i\vf} +
\sum_l a_{l} (T)  e^{ilN(T_1 - \vf)}
\label{suco}
\ee
The computation of the coefficients $a_l (T)$ is a
complicated problem \ci{progr} yet related
problems are studied in the enumerative geometry
\ci{cayley,fulton}. If we were interested in the actual
values of the correlators in the theory $\bf I$
we could use the techniques of \ci{km}. Our goal
is slightly different.

{\it Remark.} Consider  the trivial non-linear
sigma model with the target space ${\IC\IP}^{0} = $
a point. It becomes rather interesting once replaced
  by the linear gauged sigma model.
The latter has the moduli space \be
{\overline{\CM}}_{d} = {\IP}^{d} =
{\rm Sym}^{d} {\IP}^1 \label{symmi} \ee

\section{Back to future: four dimensional instantons
with freckles}

So far we have seen that adding some kind of point-like
topological defects -- ``freckles'' -- may lead to the compactification
of the moduli space of instantons in a theory {\bf I}:
${\CM} \hookrightarrow {\overline{\CM}}$.
We have also seen
that these point-like defects may be perfectly smooth
field configurations in the theory {\bf II}, but of some
characteristic size $\rho$. The natural question is:
can this work in four dimensions?

In four dimensions we study gauge theories with
instantons. The moduli space of instantons
${\CM}$
is
non-compact
due to the well-known phenomenon
of shrinking of instantons. Suppose
that the euclidean space-time is
a compact
K\"ahler surface $S$, with K\"ahler form $\omega$.
It is well-known that the moduli space
of instantons in a given gauge bundle $E$ with the
characteristic classes $c_{1}(E), c_{2}(E)$ is isomorphic
to the
moduli space of $\omega$-semistable holomorphic
bundles ${\CE}$ over $S$ with the same Chern classes
as $E$. We recall the notion of (semi)stability below.
Algebraic geometers replace the
holomorphic bundle $\CE$
by the sheaf $\CE$ of its holomorphic
sections. More specifically, over each open set $U$
one considers the abelian group
${\Gamma}\left( {\CE}\vert_{U} \right)$ of the
holomorphic sections of ${\CE}$ over $U$. The
elements $s$ of this group can be multiplied
by the holomorphic functions $f$ on $U$, $f
\in {\CO}_{U}$. This operation
makes ${\Gamma}\left({\CE}\vert_{U} \right)$
a module over the ring ${\CO}_{U}$
of holomorphic functions in $U$. For sufficiently small
$U$ one can find
a basis $s_i$ in the space of holomorphic sections of
${\CE}\vert_{U}$ such that every section $s \in
{\Gamma}\left( {\CE}\vert_{U} \right)$ can be uniquely
expanded as:
\be
s = \sum_{i=1}^{r} f_{i} s_{i}, \qquad f_{i} \in {\CO}_{U}
\label{locfr}
\ee
The sheaves with this property are called {\it locally
free} -- if the sheaf is locally free then it
is a sheaf of sections of some holomorphic bundle.
One can relax the condition \re{locfr} to the property
of being {\it torsion free}:
\be
{\rm If} \,\, f\, s = 0 \Rightarrow {\rm either}
\, f = 0 \,\, {\rm or} \,\, s = 0
\label{trfr}
\ee
The sheaf which is torsion free in general does not
come from a holomorphic bundle. Nevertheless
the beautiful property of the torsion free sheaves
(in complex dimension
two) is
that every torsion free sheaf $\CF$ is almost a bundle,
in fact over a
complement to a finite number of points in $S$
it is a bundle!  One can
always find (\ci{okonek}, lemma 1.1.8)
a holomorphic vector bundle ${\CE}$
such that ${\CE}/{\CF} = {\CS}_{Z}$,
in other words there is
an
exact sequence:  \be 0 \to \CF \to \CE \to
{\CS}_{Z} \to 0 \label{exct} \ee where ${\CS}_{Z}$
is a skyscraper sheaf
supported at points, i.e. ${\rm dim}Z = 0$,
$$ {\Gamma} \left( {\CS}_{Z}
\vert_{U} \right) = {\IC}^{\# Z \cap U } $$

We see
the first similarity between the torsion free sheaves
and vortices in two dimensions: both differ from the
honest instanton (the face of the freckled instanton)
only at finite number of
points (freckles).
More quantitatively this similarity is supported
by the fact that these points carry instanton charge:
\be
c_{2}({\CF}) = c_{2}({\CE}) - c_{2}({\CS}_{Z}) =
c_{2}({\CE}) + \# Z
\label{chrg}
\ee
The importance of the torsion free sheaves in
the studies of $S$-duality was first advocated in
\ci{avatar}. One can show that the natural backgrounds
for the higher dimensional $bc$-systems studied in \ci{clash}
are again the sheaves \ci{progr} rather then holomorphic
bundles alone.

In order for the torsion free sheaves to be useful we
need  a way to construct their moduli space
and make sure that it is compact. It turns
out that if one takes the space of
$\omega$-(semi)stable torsion free sheaves\footnote{The
sheaf ${\CF}$ is $\omega$-semistable if for
any subsheaf ${\CF}^{\prime}$ one has
$\int_{S} \o \wedge \left(
\frac{c_{1}({\CF}^{\prime})}{{\rm rk}{\CF}^{\prime}}
-
\frac{c_{1}({\CF})}{{\rm rk}{\CF}} \right) \leq 0$}
   then
the resulting space ${\overline\CM}_{c_*}$
is compact (and even
quasiprojective \ci{gieseker}).
Moreover, it can be described rather
explicitly in the case of the manifolds
$S \approx {\IP}^2$, ${\IP}^1 \times {\IP}^1$ using
ADHM techniques.

To make further contact with the two dimensional
story we need: $a)$ the supersymmetric
gauge theory whose moduli space of BPS
fields is ${\overline\CM}_{c_{*}}$, $b)$ an analogue
of the computation \re{cntr}.

The answer to the point $a)$ is not completely known.
In the light of \ci{noncomm} the natural conjecture would be
to take the theory on the ``non-commutative space''
${\hat S}$ which is a quantization of $S$
with the  $\omega^{-1}$ being the Poisson structure.

The point $b)$ is addressed using the monad
description\ci{beilinson,barth,dm,adhm}
of the moduli space of the torsion
free sheaves (a useful review is \ci{nakajima}).  For example, for $S =
{\IC\IP}^2$ the $\omega$-semistable torsion free sheaves $\CF$ arise as
follows. Let $V_{0}, V_{1}, V_{2}$ be the complex vector spaces of dimensions
$v_{0,1,2}$ respectively. Consider the
complex of bundles over $S$:
\be
0 \to V_{0} \otimes {\CO}(-1)
\longrightarrow^{\kern -10pt a} \quad
V_{1} \otimes {\CO} \longrightarrow^{\kern - 10pt b}
\quad V_{2} \otimes {\CO}(1) \to 0
\label{mnd}
\ee
In down-to-earth
terms this sequence has the following
meaning. The maps $a,b$  in the homogeneous
coordinates $(z^{0} : z^{1} : z^{2})$ are the
matrix-valued linear functions:
$a(z) = z^{\a} a_{\a}, b(z) = z^{\a} b_{\a}$. The words
``complex'' mean that
\be
b(z) \cdot a (z) = z^{\a}z^{\b} b_{\a} a_{\b} = 0
\Leftrightarrow
b_{\a} a_{\a} = 0, \, {\a} =0,1,2, \quad
b_{\a} a_{\b} + b_{\b} a_{\a} = 0, \, {\a} \neq {\b}
\label{cmpl}
\ee
For the pair $(b,a)$ of the maps between
the sheaves obeying \re{cmpl} we can define
a sheaf ${\CF}$ over $S$, whose space of sections
over an open set $U$ is
$$
{\Gamma}\left( {\CF}\vert_{U} \right) =
{\rm Ker} b (z) / {\rm Im} a (z), \quad
{\rm for} \quad
(z^{0} : z^{1} : z^{2}) \in U
$$
The space of monads is the space $M_{\rm mon}$
of triples
of matrices $a_{\b}
\in {\rm Hom} (V_{0}, V_{1})$, $b_{\a} \in
{\rm Hom}(V_{1}, V_{2})$ obeying \re{cmpl}. This space
is acted on by the group
$G_{\rm mon}^{c} = \left( {\rm GL} (V_{0} )
\times {\rm GL}(V_{1}) \times
{\rm GL}(V_{2}) \right) /{\IC}^{\star}$:
\be
(b,a) \mapsto g \cdot (b,a) = (g_{2} b g_{1}^{-1}, g_{1} a
g_{0}^{-1}),
\, {\rm for } \, (g_{0}, g_{1}, g_{2}) \in
G^{c}_{\rm mon}
\label{ge}
\ee
The sheaves defined by the pairs $(b,a)$ and
$g \cdot (b,a)$ are isomorphic.
The maximal compact subgroup of $G^{c}_{\rm mon}$
$$G_{\rm mon}
 \approx \left( U(V_0) \times U(V_1) \times
U(V_2) \right) / U(1)$$ acts in $M_{\rm mon}$
preserving its natural symplectic structure
\be
\Omega = \frac{1}{2i} \sum_{\b} {\Tr} {\d} a_{\b} \wedge
{\d} a_{\b}^{\dagger} + \frac{1}{2i} \sum_{\a} {\Tr}
{\d} b_{\a}^{\dagger} \wedge {\d} b_{\a}
\label{smpli}
\ee
Fix the real
numbers $r_{0}, r_{1}, r_{2}$,
such that
$\sum_{\a} v_{\a} r_{\a} = 0$, $r_{0}, r_{2} > 0$.
Write the moment maps:
$$
\mu_{1} = - r_{0} {\bf 1}_{v_{0}} +
\sum_{\b} a_{\b}^{\dagger} a_{\b} $$
\be
\mu_{2} = - r_{1} {\bf 1}_{v_{1}} +
\sum_{\a} b_{\a}^{\dagger} b_{\a} -
\sum_{\b} a_{\b} a_{\b}^{\dagger} \label{mmnts} \ee
$$
\mu_{3} = - r_{2} {\bf 1}_{v_{2}} +
\sum_{\a} b_{\a} b_{\a}^{\dagger}
$$
Then the moduli space of the semistable sheaves is
\be
{\overline\CM}_{c_{*}} = \left( \mu_{1}^{-1}(0) \cap
\mu_{2}^{-1}(0) \cap \mu_{3}^{-1}(0) \right) /G_{\rm mon}
\label{quot}
\ee
The compactness of the space \re{quot} is obvious:
if we first perform a reduction with respect to the groups
$U(V_{0}) \times U(V_{2})$ then the resulting space is
the product of two Grassmanians: ${\rm Gr}(v_{0}, 3v_{1})
\times {\rm Gr}(v_{2}, 3v_{1})$ which is already compact.
The subsequent reduction does not spoil this.

The Chern classes of the sheaf $\CF$ determined by the
pair $(b,a)$ are:
\be
r = v_{1} - v_{0} - v_{2}, \, c_{1} =
\left( v_{0} - v_{2} \right) \o, \, c_{2} = \frac{1}{2}
\left( \left( v_{2} - v_{0} \right)^2 + v_{0} +
v_{2} \right)
\label{chrns}
\ee
Let $(i\psi, i\phi, i\chi)$ denote the elements of the
Lie algebra of $G_{\rm mon}$, i.e.
$i\psi \in {\rm u}(V_{0}), i\phi \in {\rm u}(V_{1}),
i\chi \in {\rm u}(V_{2})$ and
$(\psi, \phi, \chi) \sim (\psi + {\bf 1}_{v_{0}},
\phi +{\bf 1}_{v_{1}}, \chi + {\bf 1}_{v_{2}})$.
We are interested in computing certain integrals over
${\overline{\CM}}_{c_{*}}$. This can be accomplished by
computing an integral over $M_{\rm mon}$ with the
insertion of the delta function in $\mu_{i}$ and
dividing by the volume of $G_{\rm mon}$ provided that the
expression we integrate is
$G_{\rm mon}$-invariant:\ci{witdgt}
\be
\int_{{\overline\CM}_{c_{*}}} \left( \ldots \right) =
\frac{1}{{\rm Vol}(G_{\rm mon})} \int_{{\rm Lie}G_{\rm mon}}
d \psi d \phi d \chi e^{i {\Tr} \psi \mu_1 +
i {\Tr} \phi \mu_2 +
i {\Tr} \chi \mu_3 } \left( \ldots \right)
\label{intgr}
\ee
The useful fact is that the observables of the
gauge theory we are interested in are the gauge-invariant
functions on $(\psi, \phi, \chi)$ only. More specifically,
there is a {\it universal sheaf} $\CU$
over
${\overline{\CM}}_{c_{*}} \times S$,
whose Chern character is
represented by:\footnote{ the universal sheaf
is defined again as ${\rm Ker}b(z)/{\rm Im} a(z)$
but now
the space of parameters contains $(b,a)$ in addition to
$z$} \be Ch ({\CU}) = {\Tr} e^{\phi} - {\Tr} e^{\psi -
\omega} - {\Tr} e^{\chi + \omega} \label{unv} \ee In
particular (cf. \re{trcs}):  $${\CO}^{(0)}_{u_{1}} =
\frac{1}{2} \left( {\Tr} {\chi}^2 +
{\Tr} {\psi}^2 - {\Tr} {\phi}^2 \right),$$
$$
\int_{S} \omega \wedge {\CO}^{(2)}_{u_{1}} =
{\Tr} {\chi} - {\Tr} {\psi}
$$
Since the observables are expressed through
$\psi, \phi, \chi$
only we can integrate out $a_{\b}, b_{\a}$ in \re{intgr}
to obtain:
$$
\langle \exp t_{1}  {\CO}^{(0)}_{u_{1}} +  T_{1} \int_{S}
\omega \wedge
{\CO}^{(2)}_{u_{1}} \rangle^{\rm torsion \, free}
=
\oint \prod_{i,j,k} d \psi_{i} d \chi_{j} d\psi_{k}
$$
$$\frac{
\prod_{i^{\prime} < i^{\prime\prime}} \left(
\psi_{i^{\prime}}
-
\psi_{i^{\prime\prime}} \right)^2
\prod_{j^{\prime} < j^{\prime\prime}} \left(
\phi_{j^{\prime}}
-
\phi_{j^{\prime\prime}} \right)^2
\prod_{k^{\prime} < k^{\prime\prime}} \left(
\chi_{k^{\prime}}
-
\chi_{k^{\prime\prime}} \right)^2
\prod_{i, k} \left( \chi_{k} - \psi_{i} \right)^6}
{\prod_{i,j} \left( \phi_{j} - \psi_{i} + i0\right)^3
\prod_{j,k} \left( \chi_{k} - \phi_{j}  + i0\right)^3 }$$
\be
\times e^{t_{1}
\frac{1}{2} \left( \sum_k {\chi_k}^2 +
\sum_i {\psi}_i^2 - \sum_j {\phi}_j^2 \right)
+ T_1 \left( \sum_k \chi_k - \sum_i \psi_i \right)
+ i r_1 \sum_i \psi_i + i r_2 \sum_j \phi_j
+ i r_3 \sum_k \chi_k}
\label{integrl}
\ee
More elaborated answer to the point $b)$ together with the
explanation of the relation of this work to the
attempts of computing ADHM integrals in \ci{dkm} will be
presented elsewhere \ci{progr}.

As a final remark notice that the freckled instantons
are present even for the gauge group $U(1)$, in which case
the moduli space ${\overline{\CM}}_{d}$ is the resolution of
singularities of the $d$-th symmetric power of the
manifold $S$:
$$
{\overline{\CM}}_{d} = {\widetilde{{\rm Sym}^{d} S}}
$$
very much like in \re{symmi}.

In both cases the freckled
instantons have only freckles, but no face.

\section{Acknowldegements}

We are grateful to
A.~Braverman,
D.~Gaitsgory,
A.~Gorsky,
M.~Finkelberg,
A.~Levin,
G.~Moore,
A.~Rosly,
E.~Witten for discussions.

The research of A.~L.~ is partly supported by  RFFI
under the grant 98-01-00328,
that of N.~N by Harvard Society of Fellows, partially by
NSF under
grant PHY-98-02709, partially by
RFFI under grant 98-01-00327; A.~L. and N.~N. are
partially supported by grant
96-15-96455 for scientific schools. Research of
S.~Sh.is supported
 by DOE grant
DE-FG02-92ER40704, by NSF CAREER award, by OJI award from
DOE and by Alfred
P.~Sloan foundation.

\end{document}